\documentclass{article}

\PassOptionsToPackage{numbers, compress}{natbib}
\usepackage{natbib}
\usepackage[final]{neurips_2023}

\usepackage[utf8]{inputenc} % allow utf-8 input
\usepackage[T1]{fontenc}    % use 8-bit T1 fonts
\usepackage{hyperref}       % hyperlinks
\usepackage{url}            % simple URL typesetting
\usepackage{booktabs}       % professional-quality tables
\usepackage{amsfonts}       % blackboard math symbols
\usepackage{amsmath}
\usepackage{nicefrac}       % compact symbols for 1/2, etc.
\usepackage{microtype}      % microtypography
\usepackage{xcolor}         % colors
\usepackage{cleveref}
\usepackage{graphicx} 

\title{Uncertainty Quantification For Turbulent Flows with Machine Learning}

\author{%
  Minghan Chu\\
  Department of Mechanical and Materials Engineering\\
  Queen's University\\
  \texttt{17MC93@queensu.ca} \\
  \And
  Weicheng Qian\\
  Department of Computer Science\\
  University of Saskatchewan\\
  \texttt{weicheng.qian@usask.ca} \\
}

\begin{document}

\maketitle

\begin{abstract}
  Turbulent flows are of central importance across applications in science and engineering problems. For design and analysis, scientists and engineers use Computational Fluid Dynamics (CFD) simulations using turbulence models. Turbulent models are limited approximations, introducing epistemic uncertainty in CFD results. For reliable design and analysis, we require quantification of these uncertainties. The Eigenspace Perturbation Method (EPM) is the preeminent physics based approach for turbulence model UQ, but often leads to overly conservative uncertainty bounds. In this study, we use Machine Learning (ML) models to moderate the EPM perturbations and introduce our physics constrained machine learning framework for turbulence model UQ. We test this framework in multiple problems to show that it leads to improved calibration of the uncertainty estimates. 
\end{abstract}

\section{Overview}
Turbulent fluid flows are of central importance across problems in Science and Engineering. For example, the design of transcatheter aortic valves (TAV) focuses on minimizing turbulence in the blood flow for patient safety\cite{pietrasanta2022characterization}. The design of automobiles aims to reduce turbulence in the automobile wake to save energy\cite{pope2001turbulent}. Such design studies use Computational Fluid Dynamics (CFD) simulations with turbulence models to account for the effects of turbulence. Turbulence models are simple constitutive equations that relate the effects of turbulence to measurable variables. This simplification is an advantage for computational expense, but also leads to severe limitation in the turbulence physics that these models can replicate. This limitation leads to epistemic uncertainty in CFD predictions, which can have hazardous ramifications on engineering designs. Estimating these uncertainties is essential for reliable design\cite{oliver2009uncertainty, zeng2022parametric, dow2011uncertainty, stephanopoulos2016uncertainty}. The only physics based approach to estimate turbulence model uncertainty is the Eigenspace Perturbation Method (EPM)\cite{iaccarino2017eigenspace}. This uses physics based perturbations in the turbulence model predictions to estimate predictive uncertainty. Studies of turbulence model Uncertainty Quantification (UQ) are mainly focused on the EPM based eigenvalue and eigenvector perturbations, such as turbulent flow through scramjets \cite{emory2011characterizing}, aircraft nozzle jets \cite{mishra2017uncertainty}, over steamlined bodies \cite{gorle2019epistemic}, in aeronautical engines\cite{cook2019optimization, mishra2020design}, supersonic axisymmetric submerged jet \cite{mishra2017rans}, canonical cases of turbulent flows over a backward-facing step \cite{iaccarino2017eigenspace,cremades2019reynolds}, and benchmark cases of complex turbulent flow \cite{thompson2019eigenvector}. Its theoretical foundations are well established\cite{mishra2019theoretical} and the EPM's software implementations are widely used\cite{mishra2019uncertainty}. Despite its success, the EPM has limitations. Primarily, the EPM weighs all physically permissible events equally, leading to uncalibrated and conservative prediction intervals. This leads to overly safe and inefficient designs. If we can weigh all physically permissible events by their likelihood, we can improve the calibration of the uncertainty bounds. In this study, we augment the EPM framework with Machine Learning models to infer the strength of the perturbations for better calibration of uncertainty bounds.   

\section{Methods \& Methodology}
In this study, we use the turbulence model of Langtry and Menter \cite{langtry2009correlation} to simulate turbulent flow over an SD7003 airfoil at $8^\circ$ angle of attack (AoA). With the Reynolds number based on the cord length of $Re_{c} = 60000$, the flow underwent transition to turbulence on the suction side of the airfoil. Hereon, the turbulence model predictions for this airfoil case are referred to as RANS (Reynolds Averaged Navier Stokes) and the true targets as DNS (Direct Numerical Simulation). Such turbulent flows are characterised by random fluctuations in the velocity and pressure fields, where the instantaneous velocity can be decomposed into a mean and fluctuating component, $u=U_{mean}+u_{fluctuation}$. The key quantity of interest for design is the covariance of this fluctuating velocity referred to as the Reynolds Stress Tensor, $\left\langle u_i u_j\right\rangle$, that encompasses the effect of turbulence on the flow. The trace of the Reynolds Stress Tensor is referred to as the turbulence kinetic energy, $k$. In the EPM\cite{iaccarino2017eigenspace}, the perturbed Reynolds stresses are defined as
\begin{equation}\label{Eq:Rij_perturb}
        \left\langle u_{i} u_{j}\right\rangle^{*}=2 k^{*}\left(\frac{1}{3} \delta_{i j}+v_{i n}^{*} \hat{b}_{n l}^{*} v_{j l}^{*}\right),
\end{equation}
where $k^{*}$ is the perturbed turbulence kinetic energy, $\hat{b}_{k l}^{*}$ is the perturbed eigenvalue matrix for the Reynolds Stress Tensor, $v_{i j}^{*}$ is perturbed eigenvector matrix, and $\delta_{i j}$ is the kronecker delta. In this study the perturbed turbulence kinetic energy can be defined as
\begin{equation}\label{Eq:Marker_Mk_Method}
    k^{*} = k +\Delta_k = M_{k}, \quad M_{k} \sim f(x,y),
\end{equation}
where $M_{k}$ is a marker function of the $x$ and $y$ coordinate in a computational domain. This marker function predicts the magnitude of the perturbation to be used to modulate EPM perturbations. 

In this study, we examined polynomial regression to construct this marker function augmented with eigenvalue perturbations to estimate the uncertainty bound for the predicted skin friction coefficient. We also train a convolutional neural network (CNN) to predict high-fidelity turbulence kinetic energy. Here, we use a one-dimensional convolutional neural network to learn the projection from the functional mapping estimated by the turbulence model $f^{\scriptstyle\mathrm{RANS}}(x, y)$ to the true mapping $f^{\mathrm{DNS}}(x, y)$. For a given $x$, we can rewrite the estimated function as $g_x(k,  y)$. Assuming that there exists a morphism $F$ from $g^{\mathrm{RANS}}_x(k,  y)$ to $g^{\mathrm{DNS}}_x(k,  y)$, every $x$ and $g_x(k, y)$ is smooth. Our CNN is trained to depict $F$ so as to project $g^{\mathrm{RANS}}_x(k,  y)$ to $g^{\mathrm{DNS}}_x(k,  y)$. This is conducted by training the paired RANS- and DNS-estimated functions at the selected $x$ coordinates. Taking advantage of the smoothness assumption of $g_x(k, y)$, our 1D-CNN is trained to predict DNS estimated function at $(x, y_\mathrm{target})$ given a series of RANS estimated function at $(x, y_i)$, where $y_i \in [y - \epsilon, y + \epsilon], \epsilon > 0$ belongs to the neighbor of $y_\mathrm{target}$. Our 1D-CNN has four-layers and in total 86 parameters: a single model for all zones at any $x$ to project RANS to DNS. We trained our 1D-CNN with normalized pairs of $\left( g^{\mathrm{RANS}}_x(k,  y), g^{\mathrm{DNS}}_x(k,  y) \right)$ at only three positions $x = 0.4, 0.56, 0.58$ with mean squared error as the loss function and a 80\%--20\% split as training--testing dataset. We validated our trained 1D-CNN by comparing the L1 loss of RANS, denoted as $L^1_c(\texttt{rans}) = \lvert CF^{\mathrm{RANS}}_{k} - CF^{\mathrm{DNS}}_{k} \rvert$ with the L1 loss of 1D-CNN projected RANS, denoted as $L^1_c(\texttt{pred}) = \lvert CF^{\mathrm{CNN}}_{k} - CF^{\mathrm{DNS}}_{k} \rvert$.

\begin{figure}[h!]
%outer boundary layer     
         \centering
         \includegraphics[width=8cm]{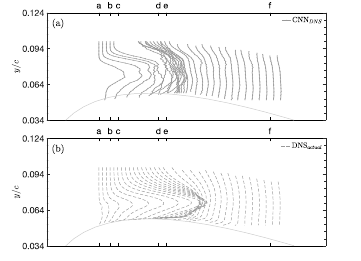}
        \caption{CNN projected DNS (\texttt{CNN\_{DNS}}) compared with ground truth (\texttt{DNS\_{actual}}). There are 32 positions on the suction side of the airfoil.}
        \label{fig:CNN_DNS.pdf}
\end{figure}

\section{Results}
\begin{figure}[h!]
%outer boundary layer     
         \centering
         \includegraphics[width=12cm]{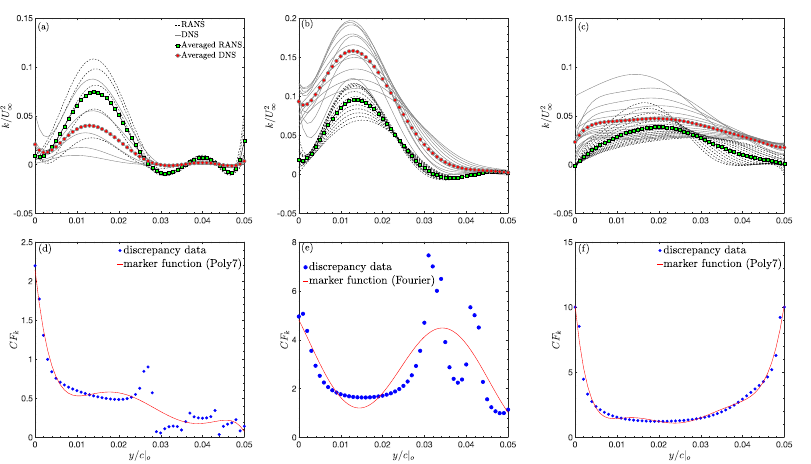}
        \caption{Mean of $7^{th}$ order polynomials for normalized turbulence kinetic energy (a) - (c), and the corresponding marker function (d) - (f). (a) and (d) zone $ab$; (b) and (e) zone $cd$; (c) and (f) zone $ef$.}
        \label{fig:Discrepancy_Marker.pdf}
\end{figure}

\begin{figure}[h!]
%outer boundary layer     
         \centering
         \includegraphics[width=6cm]{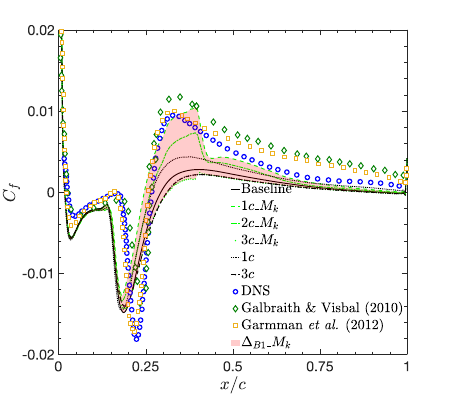}
        \caption{Skin friction coefficient. Displayed are uncertainty bounds for $1c\_M_{k}$, $2c\_M_{k}$ and $3c\_M_{k}$ perturbations (red envelope). $\Delta_{B1}$ stands for $\Delta_{B} = 1$. Profile of the baseline prediction and eigenvalue perturbations ($1c$ and $3c$) are provided for reference.}
        \label{fig:Cf.pdf}
\end{figure}

\begin{figure}[h!]
%outer boundary layer     
    \centering
    \includegraphics[width=\textwidth]{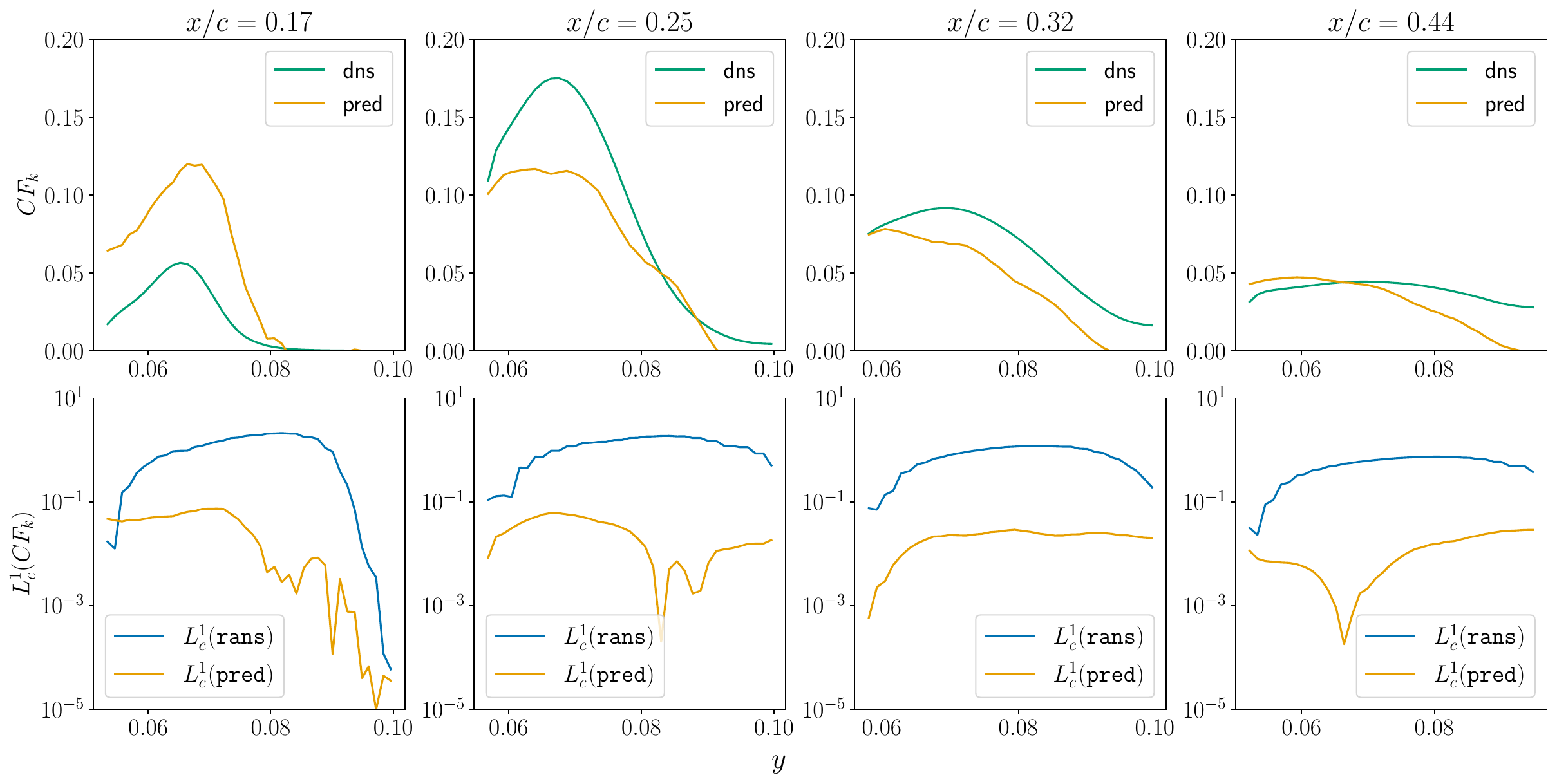}
    \caption{First row: CNN projected DNS (\texttt{pred}) compared with ground truth (\texttt{dns}). Second row: Validation of 1D-CNN by comparing L1 loss between $L^1_c(\texttt{rans})$ and $L^1_c(\texttt{pred})$.}
    \label{fig:cnn-projected-dns-with-rands.pdf}
\end{figure}

In Figures \ref{fig:Discrepancy_Marker.pdf} (a) - (c), the mean of the polynomial regression-based normalized turbulence kinetic energy profile as the representative for each zone shows the discrepancy between RANS and DNS. Note that all profiles are shifted down to the origin of $y/c$, denoted $y /\left.c\right|_o$. The corresponding marker for each zone is shown in Figs. \ref{fig:Discrepancy_Marker.pdf} (d) - (f). Marker functions can be constructed by fitting appropriate models to the discrepancy data for each zone, i.e., a seventh-order polynomial for the $ab$ and $ef$ zone, and a Fourier series for the $cd$ zone. Augmenting the EPM eigenvalue perturbation \footnote{The strength of eigenvalue perturbation is denoted $\Delta_{B}$, which varies from $0$ to $1$.} with the marker function ($1c\_M_{k}$, $2c\_M_{k}$ and $3c\_M_{k}$) using Eqs. \ref{Eq:Rij_perturb} and \ref{Eq:Marker_Mk_Method}, the estimated model-form uncertainty (red envelope) for the predicted skin friction coefficient is constructed and shown in Fig. \ref{fig:Cf.pdf}. The $1c$ and $3c$ eigenvalue perturbations are included for reference. It is clear that the uncertainty bound successfully encompasses the ILES/LES data of \cite{galbraith2010implicit} and \cite{garmann2013comparative} for $0.25 < x/c < 0.45$. This region falls into the $cd$ and part of the $ef$ zone, where the separation bubble is forming and the flow is re-attaching on the wall surface. In comparison to the eigenvalue perturbations, the red envelop exhibits a significant increase in the magnitude of $C_{f}$, exhibiting a tendency to retain the shape of the reference data. This marks a significant improvement in the RANS model prediction for $C_{f}$. The shape of the red envelope is not as smooth as the eigenvalue perturbations, reflecting the effect of spatial variability in $M_{k}$. The 1D-CNN can predict DNS at any zone given RANS, thus acting as the marker function $M_{k}$ in Eq. \ref{Eq:Marker_Mk_Method}. From Figs. \ref{fig:CNN_DNS.pdf} (a) and (b), the CNN predicted DNS profile for $k$ exhibits agreement with the DNS dataset. In Fig. \ref{fig:cnn-projected-dns-with-rands.pdf}, the series of CNN predicted DNS profiles in the first row are then smoothed with the moving average with a window size of six consecutive estimations. Our CNN predicted DNS profiles resemble the ground truth DNS despite being trained with only a few pairs of RANS and DNS results. From Fig. \ref{fig:cnn-projected-dns-with-rands.pdf}, the discrepancy in general reduces as the flow proceeds further downstream. Consequently, the CNN predicted DNS given the RANS estimated function acts as the marker function $M_k$ in Eq. \ref{Eq:Marker_Mk_Method}. From the Fig. \ref{fig:cnn-projected-dns-with-rands.pdf}, the second row shows the computed error of the baseline solution and the CNN predicted DNS, and it is clear that the error for CNN predicted DNS is significantly reduced in magnitude compared to that for the baseline solution. 

\section{Conclusions}
\label{sec:Conclusion}

We investigated if ML models, specifically polynomial regression and CNNs, can augment the Eigenspace Perturbation Framework to give better calibration of uncertainty intervals.  The learning algorithms need to be coupled to the EPM implemented in OpenFOAM\cite{weller1998tensorial} to construct a marker function for the turbulence kinetic energy perturbation. ML models capture the discrepancy in the predicted turbulence kinetic energy between RANS and DNS. Correspondingly, the marker function is augmented with the  eigenvalue perturbation to significantly increase the uncertainty bound for $C_{f}$. Around the peak of the $C_{f}$ curve, the uncertainty bound successfully encompassed separation bubble. While researchers have attempted to use ML models to augment the EPM\cite{heyse2021data, heyse2021estimating, matha2023evaluation}, we are the first to examine the projection from RANS to DNS using the CNN approach. Our experiment results suggest that the CNN approach can help us project the RANS estimated marker function to DNS data. A projection that can approximate the DNS reasonably well from RANS might exist independent of $x$. Our methodology can be easily extended to analyze flows over different airfoils. Future work may include evaluating other machine learning models in generating marker functions with different types of airfoils, as well as integrating the CNN approach into the EPM.

\section{Impact statement}
Transitional flows are frequently encountered in aerospace and medical applications. While turbulence models have severe challenges in such flows, they are the only pragmatic recourse. Thus the estimation of turbulence model uncertainty is valuable for improving the usefulness of turbulence models in engineering applications. A recent physics-based eigenspace perturbation method evaluates the accuracy of turbulence models for practical usage based on physically possible perturbations, which is less reliable in terms of giving exact strength of perturbation. We propose a machine learning augmented eigenspace perturbation method that can effectively increase the precision of the estimates of turbulence model uncertainty and build confidence in engineering simulations. Our method is generalizable to a variety of flow scenarios. Our method can also be employed to shed lights on predicting errors in turbulence model predictions, enabling a correction to turbulence models to improve their accuracy.

\bibliography{bib}

\begin{thebibliography}{10}

\bibitem{pietrasanta2022characterization}
Leonardo Pietrasanta, Shaokai Zheng, Dario De~Marinis, David Hasler, and Dominik Obrist.
\newblock Characterization of turbulent flow behind a transcatheter aortic valve in different implantation positions.
\newblock {\em Frontiers in cardiovascular medicine}, 8:804565, 2022.

\bibitem{pope2001turbulent}
Stephen~B Pope.
\newblock Turbulent flows, 2001.

\bibitem{oliver2009uncertainty}
Todd Oliver and Robert Moser.
\newblock Uncertainty quantification for rans turbulence model predictions.
\newblock In {\em APS division of fluid dynamics meeting abstracts}, volume~62, pages LC--004, 2009.

\bibitem{zeng2022parametric}
Fan-zhi Zeng, Jin-ping Li, Yu~Wang, Mao Sun, and Chao Yan.
\newblock Parametric uncertainty quantification of sst turbulence model for a shock train and pseudo-shock phenomenon.
\newblock {\em Acta Astronautica}, 196:290--302, 2022.

\bibitem{dow2011uncertainty}
Eric Dow and Qiqi Wang.
\newblock Uncertainty quantification of structural uncertainties in rans simulations of complex flows.
\newblock In {\em 20th AIAA Computational Fluid Dynamics Conference}, page 3865, 2011.

\bibitem{stephanopoulos2016uncertainty}
Kimon Stephanopoulos, Isaac Witte, Tim Wray, and Ramesh~K Agarwal.
\newblock Uncertainty quantification of turbulence model coefficients in openfoam and fluent for mildly separated flows.
\newblock In {\em 46th AIAA Fluid Dynamics Conference}, page 4401, 2016.

\bibitem{iaccarino2017eigenspace}
Gianluca Iaccarino, Aashwin~Ananda Mishra, and Saman Ghili.
\newblock Eigenspace perturbations for uncertainty estimation of single-point turbulence closures.
\newblock {\em Physical Review Fluids}, 2(2):024605, 2017.

\bibitem{emory2011characterizing}
Michael Emory, Vincent Terrapon, Rene Pecnik, and Gianluca Iaccarino.
\newblock Characterizing the operability limits of the hyshot ii scramjet through rans simulations.
\newblock In {\em 17th AIAA international space planes and hypersonic systems and technologies conference}, page 2282, 2011.

\bibitem{mishra2017uncertainty}
Aashwin~Ananda Mishra and Gianluca Iaccarino.
\newblock Uncertainty estimation for reynolds-averaged navier--stokes predictions of high-speed aircraft nozzle jets.
\newblock {\em AIAA Journal}, 55(11):3999--4004, 2017.

\bibitem{gorle2019epistemic}
C~Gorl{\'e}, S~Zeoli, M~Emory, J~Larsson, and G~Iaccarino.
\newblock Epistemic uncertainty quantification for reynolds-averaged navier-stokes modeling of separated flows over streamlined surfaces.
\newblock {\em Physics of Fluids}, 31(3):035101, 2019.

\bibitem{cook2019optimization}
Laurence~W Cook, AA~Mishra, JP~Jarrett, KE~Willcox, and G~Iaccarino.
\newblock Optimization under turbulence model uncertainty for aerospace design.
\newblock {\em Physics of Fluids}, 31(10):105111, 2019.

\bibitem{mishra2020design}
Aashwin~Ananda Mishra, Jayant Mukhopadhaya, Juan Alonso, and Gianluca Iaccarino.
\newblock Design exploration and optimization under uncertainty.
\newblock {\em Physics of Fluids}, 32(8):085106, 2020.

\bibitem{mishra2017rans}
AA~Mishra and G~Iaccarino.
\newblock Rans predictions for high-speed flows using enveloping models.
\newblock {\em arXiv preprint arXiv:1704.01699}, 2017.

\bibitem{cremades2019reynolds}
Luis~F Cremades~Rey, Denis~F Hinz, and Mahdi Abkar.
\newblock Reynolds stress perturbation for epistemic uncertainty quantification of rans models implemented in openfoam.
\newblock {\em Fluids}, 4(2):113, 2019.

\bibitem{thompson2019eigenvector}
Roney~L Thompson, Aashwin~Ananda Mishra, Gianluca Iaccarino, Wouter Edeling, and Luiz Sampaio.
\newblock Eigenvector perturbation methodology for uncertainty quantification of turbulence models.
\newblock {\em Physical Review Fluids}, 4(4):044603, 2019.

\bibitem{mishra2019theoretical}
AA~Mishra and G~Iaccarino.
\newblock Theoretical analysis of tensor perturbations for uncertainty quantification of reynolds averaged and subgrid scale closures.
\newblock {\em Physics of Fluids}, 31(7):075101, 2019.

\bibitem{mishra2019uncertainty}
Aashwin~Ananda Mishra, Jayant Mukhopadhaya, Gianluca Iaccarino, and Juan Alonso.
\newblock Uncertainty estimation module for turbulence model predictions in su2.
\newblock {\em AIAA Journal}, 57(3):1066--1077, 2019.

\bibitem{langtry2009correlation}
Robin~B Langtry and Florian~R Menter.
\newblock Correlation-based transition modeling for unstructured parallelized computational fluid dynamics codes.
\newblock {\em AIAA journal}, 47(12):2894--2906, 2009.

\bibitem{galbraith2010implicit}
Marshall Galbraith and Miguel Visbal.
\newblock Implicit large eddy simulation of low-reynolds-number transitional flow past the sd7003 airfoil.
\newblock In {\em 40th fluid dynamics conference and exhibit}, page 4737, 2010.

\bibitem{garmann2013comparative}
Daniel~J Garmann, Miguel~R Visbal, and Paul~D Orkwis.
\newblock Comparative study of implicit and subgrid-scale model large-eddy simulation techniques for low-reynolds number airfoil applications.
\newblock {\em International Journal for Numerical Methods in Fluids}, 71(12):1546--1565, 2013.

\bibitem{weller1998tensorial}
Henry~G Weller, Gavin Tabor, Hrvoje Jasak, and Christer Fureby.
\newblock A tensorial approach to computational continuum mechanics using object-oriented techniques.
\newblock {\em Computers in physics}, 12(6):620--631, 1998.

\bibitem{heyse2021data}
Jan~Felix Heyse, Aashwin~Ananda Mishra, and Gianluca Iaccarino.
\newblock Data driven physics constrained perturbations for turbulence model uncertainty estimation.
\newblock In {\em AAAI Spring Symposium: MLPS}, 2021.

\bibitem{heyse2021estimating}
Jan~Felix Heyse, Aashwin~A Mishra, and Gianluca Iaccarino.
\newblock Estimating rans model uncertainty using machine learning.
\newblock {\em Journal of the Global Power and Propulsion Society}, 2021(May):1--14, 2021.

\bibitem{matha2023evaluation}
Marcel Matha, Karsten Kucharczyk, and Christian Morsbach.
\newblock Evaluation of physics constrained data-driven methods for turbulence model uncertainty quantification.
\newblock {\em Computers \& Fluids}, 255:105837, 2023.

\end{thebibliography}
\appendix

\end{document}